\documentclass[sigconf]{acmart}

\AtBeginDocument{%
  \providecommand\BibTeX{{%
    \normalfont B\kern-0.5em{\scshape i\kern-0.25em b}\kern-0.8em\TeX}}}

\setcopyright{rightsretained} 
\copyrightyear{2022} 
\acmYear{2022} 
\acmConference{SIGGRAPH '22 Talks}{August 07-11, 2022}{Vancouver, BC, Canada}
\acmBooktitle{Special Interest Group on Computer Graphics and Interactive Techniques Conference Talks (SIGGRAPH '22 Talks), August 07-11, 2022}
\acmDOI{10.1145/3532836.3536263}
\acmISBN{978-1-4503-9371-3/22/08}

\usepackage{units}
\usepackage{wrapfig}

\begingroup

\begin{document}

\title{Countering Racial Bias in Computer Graphics Research}

\author{Theodore Kim}
\author{Holly Rushmeier}
\author{Julie Dorsey}
\email{theodore.kim@yale.edu}
\email{holly.rushmeier@yale.edu}
\email{julie.dorsey@yale.edu}
\affiliation{%
  \institution{Yale University}
\country{USA}
}
\author{Derek Nowrouzezahrai}
\email{derek@cim.mcgill.ca}
\affiliation{%
  \institution{McGill University}
\country{Canada}
}
\author{Raqi Syed}
\email{raqi.syed@vuw.ac.nz}
\affiliation{%
  \institution{Victoria University of Wellington}
\country{New Zealand}
}
\author{Wojciech Jarosz}
\email{wj@dartmouth.edu}
\affiliation{%
  \institution{Dartmouth College}
\country{USA}
}
\author{A.M.~Darke}
\email{darke@ucsc.edu}
\affiliation{%
  \institution{University of California, Santa Cruz}
\country{USA}
}

\renewcommand{\shortauthors}{Kim, et al.}

\maketitle

\section{Introduction}
The murder of George Floyd and the worldwide protests that erupted in its wake have foregrounded the pervasive nature of systemic racism. 
Graphics research is no exception, as racial homogeneities in the historical composition of our community have contributed to racially biased practices. The pale skin and straight hair targeted by our algorithms for virtual humans directly reflect the European and East Asian researchers that designed them \cite{KimSciAm20}.

To build a better future, and realize ACM SIGGRAPH's Vision Statement of {\em Enabling Everyone to Tell Their Stories}, we must expand our palette of research problems to encompass the full spectrum of humanity.
In the following, we will detail how racial bias pervades the technical language and numerical measures we use in research. To push against this historical inertia, we propose a {\em quaternion quarter circle} numerical measure, propose qualitative improvements to current research practices, and pose several historically-neglected research questions. Implementing these practices and investigating these questions are a first step towards a more comprehensive approach to computer graphics research.




\subsection{{\em Acts}, Not {\em People}}

For the current discussion, we define {\em acts}, and not {\em people}, as racist. In the absence of constant vigilance, we are all capable of committing {\em racist acts} that perpetuate existing systemic inequalities \cite{kendi2019}.
This perspective allows us to examine how seemingly neutral practices in computer graphics research have resulted, independent of any individual intent, in measurably biased outcomes. Our supplement provides further details and a bibliography.

Translucency and the corresponding physical mechanism of subsurface scattering has become synonymous with ``human skin'' in rendering. However, translucency is only the dominant visual feature of young, white Europeans and fair-skinned East Asians. We found 19 graphics publications, including foundational works on the topic, that solely present renderings of white humans as evidence that subsurface scattering algorithms can faithfully depict ``skin'', ``human skin'' and ``human faces.'' In at least 4 instances, this bias is then reflected in commercial software. Several other publications that include darker skin present them as deviations from the white baseline, further reinforcing the supremacy of whiteness. Researchers performing the seemingly neutral act of capturing their own appearances have instead perpetuated existing inequalities.



Similarly, ``hair'' has become synonymous with straight or wavy hair, and simulation and rendering papers cluster around this type. 
However, over a billion humans in Africa and its attendant diaspora have ``afro-textured'' or ``kinky'' hair. We only found {\em two} works in the graphics literature that attempt to capture the visual phenomena associated with these billion people. In contrast, 41 graphics publications, again including foundational works, solely present images of straight or wavy hair as evidence that the algorithms can faithfully depict ``human hair''. If we do not actively guard against our own biases, we will reproduce existing inequalities.

\subsection{Existing Quantitative Measures}

One potential solution is to use medical and cosmetics scales to quantify which human characteristics that specific graphics algorithms are intended to depict. Multiple dermatological systems measure the darkness of skin, such as the \citet{fitzpatrick1975} scale which classifies white European skin as Type I, and darker skin using progressively higher numbers up to Type VI, or the von Luschan and Taylor Hyperpigmentation scales, which assign Type 0 to white skin and respectively Types 36 and 10 to dark skin.

These measurement systems all share a problem clearly identified in Feminist and Critical Race Studies \cite{Dyer97,Phillips20}: European features are granted primary numerical status in the Fitzpatrick scale (literally the \#1 ranking), or placed at the {\em center} or {\em origin} \cite{Benjamin2019} in the cases of the von Luschan and Taylor scales. This grants white, European appearances centrality, while darker skin is placed at inconsistent locations (VI, 36, and 10) along an unbounded number line. 


Existing hair typing systems share the same problem. The Walker system widely used by stylists classifies straight hair as Type 1a. Progressively curlier hair is assigned higher numbers, up to Type 4c. The L'Or\'eal \cite{de2007shape} system assigns Type I to straight hair, and Type VIII to kinky hair. Again, European features are \#1 while Black features are classed as either 4c or VIII.



This seemingly subtle ranking has visible impacts in the graphics literature on human appearance. Past works have included Fitzpatrick Types III and IV \cite{donner2008layered} in their measurements, but final renderings still only showcase white skin. 
Other works \cite{Khungurn17} have measured hair types with different elliptical cross sections (again, European hair scores closest to 1), but the final renders only depict straight or wavy blonde hair. The implicit ranking explicitly reproduces bias.


\subsection{Proposed Anti-Racist Practices}



Adopting scales from medicine and cosmetics can seem objective, but medical practices are equally susceptible to systemic bias \cite{Sjoding20}. Moreover, borrowing an existing scale can backfire if it misaligns with the appearance space of interest. Good-faith attempts to explore skin appearance across Fitzpatrick values \cite{Weyrich2006} becomes beholden to a scale that was designed to classify skin according to its damage susceptibility under UV light and various treatments. We instead advocate multi-dimensional scales designed for visual appearance, such as the Pantone-based scale inspired by Angélica Dass \cite{carrington2020angelica}.


One-dimensional scales like Walker are appealing because they are straightforward to explore, but reproducing the range of human appearance will almost certainly require higher-dimensions. Rather than using the Fitzpatrick scale as a proxy for melanin concentration, we advocate direct reporting of concentrations, while underscoring that other factors like spatial and directional variation will introduce further non-linearities. Similarly, \citet{de2007shape} uses PCA to reduce the wavelength, curvature and helicity of hairs into a single scale, but a comprehensive visual approach will need to disentangle these factors. Flattenings are a useful way to understand {\em outputs}, but can be misleadingly simplify {\em inputs}.

Measuring and exploring the applicability of existing methods in higher-dimensional parameter spaces is an exacting and necessary process. In order to make progress, we as researchers must commit to encouraging and favorably reviewing work that both evaluates current methodologies, and develops novel techniques for measuring the applicable ranges of new and existing methods.

\setlength{\intextsep}{5pt}%
\setlength{\columnsep}{10pt}%
\begin{wrapfigure}{r}{0.15\textwidth} 
    \includegraphics[width=0.15\textwidth]{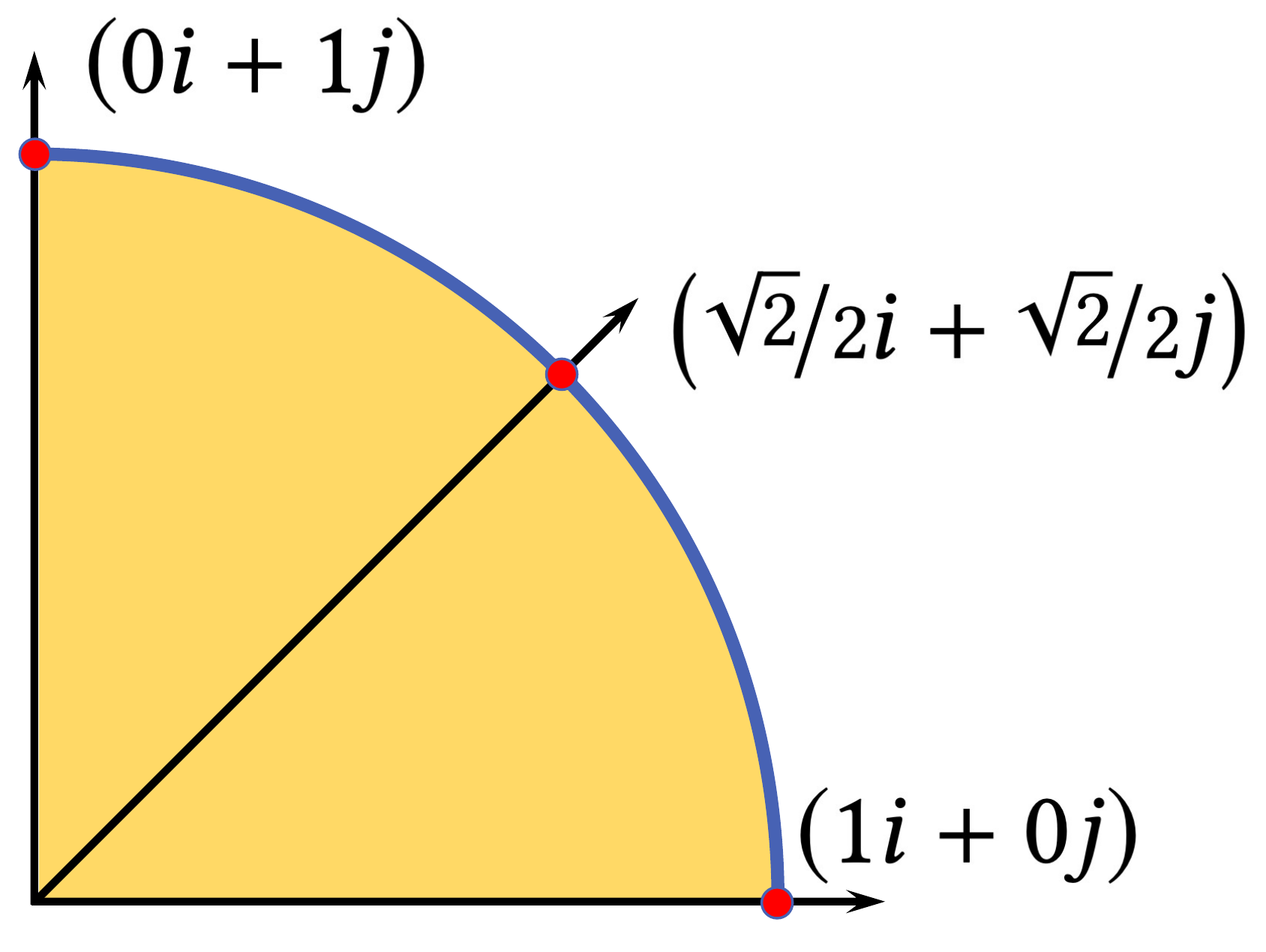}
\end{wrapfigure}
Once measurements are in hand, we must take care to avoid past mistakes, and choose a system that does not implicitly center certain populations. For this purpose, we propose a {\em quarter circle quaternion} where the real and $k$ coordinate are pinned to zero. 
This measure has several features.
\begin{itemize}
\item The origin is excluded, so no value claims ``the center.''
\item The appearance of a 1 signals the {\em endpoint} of a scale, but neither $(1i + 0j)$ nor $(0i + 1j)$ can claim the ranking of \#1.
\item All points on the scale have a numerical norm of 1. No value has a greater or lesser magnitude than any other.
\item The dueling nomenclature of {\em complex} vs.~{\em imaginary} reflects the larger social, political and biological discourse. Some will view this mapping as addressing an imaginary problem. We assert that the problems are both real and complex.
\end{itemize}
The main purpose of our quarter circle quaternion is to {\em initiate discussion of a thorny problem}, not to claim to have formulated an authoritative solution. Alternatives are possible, and welcome. 

\subsection{Future Research Directions}

The preceding discussion suggests a variety of directions. 
\begin{itemize}
\item What is a complete rendering model for Black skin? Where do the blue tones come from, and could it benefit from a custom, multi-layer BSDF model?
\item What is an efficient simulation model for kinky hair? Collisions behave differently from the straight case.
\item What is an efficient rendering model for kinky hair? Can the near-isotropic strand distributions be leveraged? Do wave-based silhouette phenomena gain importance?
\item What is the {\em gamut} of human appearance? How many coordinates are needed, and is it computationally tractable? 
\end{itemize}
When investigating these questions, it is imperative to actively recruit and engage stakeholders that both inhabit the forms of humanity being modeled, and are the intended users of the technology \cite{friedman2008value}. This is scientifically sound: nobody is better positioned to formulate algorithms that capture the subtle qualities of Black skin and kinky hair than somebody who sees them in the mirror every morning. Such practice would increase the diversity of our community and reduce the risk of publishing well-intentioned but counter-productive ``Blackface'' skin or ``Giant Afro'' hair papers. 

This work will require grappling with uncomfortable truths, and will not be immune from missteps and failure. However, these are necessary steps towards creating an environment that values self-reflection and open dialog. Progress is urgently needed, because  graphics algorithms are increasingly being used to generate training sets for machine learning, so their potential to amplify existing inequalities has never been greater \cite{Ben:Geb:McM:21b}. By learning from past missteps, we can instead move towards a future where everyone can tell their stories. We commit to rendering this future.

\bibliographystyle{ACM-Reference-Format}
\bibliography{unbiased,holly-arm}


\begin{thebibliography}{14}


\ifx \showCODEN    \undefined \def \showCODEN     #1{\unskip}     \fi
\ifx \showDOI      \undefined \def \showDOI       #1{#1}\fi
\ifx \showISBNx    \undefined \def \showISBNx     #1{\unskip}     \fi
\ifx \showISBNxiii \undefined \def \showISBNxiii  #1{\unskip}     \fi
\ifx \showISSN     \undefined \def \showISSN      #1{\unskip}     \fi
\ifx \showLCCN     \undefined \def \showLCCN      #1{\unskip}     \fi
\ifx \shownote     \undefined \def \shownote      #1{#1}          \fi
\ifx \showarticletitle \undefined \def \showarticletitle #1{#1}   \fi
\ifx \showURL      \undefined \def \showURL       {\relax}        \fi
\providecommand\bibfield[2]{#2}
\providecommand\bibinfo[2]{#2}
\providecommand\natexlab[1]{#1}
\providecommand\showeprint[2][]{arXiv:#2}

\bibitem[\protect\citeauthoryear{Bender, Gebru, McMillan-Major, and
  Mitchell}{Bender et~al\mbox{.}}{2021}]%
        {Ben:Geb:McM:21b}
\bibfield{author}{\bibinfo{person}{E. Bender}, \bibinfo{person}{T. Gebru},
  \bibinfo{person}{A. McMillan-Major}, {and} \bibinfo{person}{M. Mitchell}.}
  \bibinfo{year}{2021}\natexlab{}.
\newblock \showarticletitle{On the Dangers of Stochastic Parrots: Can Language
  Models Be Too
  Big?\raisebox{-5pt}{\includegraphics[scale=0.075]{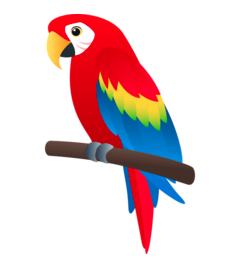}}}. In
  \bibinfo{booktitle}{\emph{Proc. of FAccT}}.
\newblock


\bibitem[\protect\citeauthoryear{Benjamin}{Benjamin}{2019}]%
        {Benjamin2019}
\bibfield{author}{\bibinfo{person}{R. Benjamin}.}
  \bibinfo{year}{2019}\natexlab{}.
\newblock \bibinfo{booktitle}{\emph{Race After Technology}}.
\newblock \bibinfo{publisher}{Polity Press}.
\newblock


\bibitem[\protect\citeauthoryear{Carrington, Vatanchi, and Jakus}{Carrington
  et~al\mbox{.}}{2020}]%
        {carrington2020angelica}
\bibfield{author}{\bibinfo{person}{A.E. Carrington}, \bibinfo{person}{M.
  Vatanchi}, {and} \bibinfo{person}{J. Jakus}.}
  \bibinfo{year}{2020}\natexlab{}.
\newblock \showarticletitle{Ang{\'e}lica Dass’ {H}umanae, a spectrum of skin
  tones}.
\newblock \bibinfo{journal}{\emph{Int. J. Dermatol.}} \bibinfo{volume}{59},
  \bibinfo{number}{5} (\bibinfo{year}{2020}), \bibinfo{pages}{640--642}.
\newblock


\bibitem[\protect\citeauthoryear{De~La~Mettrie, Saint-L{\'e}ger, Loussouarn,
  Garcel, Porter, and Langaney}{De~La~Mettrie et~al\mbox{.}}{2007}]%
        {de2007shape}
\bibfield{author}{\bibinfo{person}{R. De~La~Mettrie}, \bibinfo{person}{D.
  Saint-L{\'e}ger}, \bibinfo{person}{G. Loussouarn}, \bibinfo{person}{A.
  Garcel}, \bibinfo{person}{C. Porter}, {and} \bibinfo{person}{A. Langaney}.}
  \bibinfo{year}{2007}\natexlab{}.
\newblock \showarticletitle{Shape variability and classification of human hair:
  a worldwide approach}.
\newblock \bibinfo{journal}{\emph{Human biology}} \bibinfo{volume}{79},
  \bibinfo{number}{3} (\bibinfo{year}{2007}), \bibinfo{pages}{265--281}.
\newblock


\bibitem[\protect\citeauthoryear{Donner, Weyrich, d'Eon, Ramamoorthi, and
  Rusinkiewicz}{Donner et~al\mbox{.}}{2008}]%
        {donner2008layered}
\bibfield{author}{\bibinfo{person}{C. Donner}, \bibinfo{person}{T. Weyrich},
  \bibinfo{person}{E. d'Eon}, \bibinfo{person}{R. Ramamoorthi}, {and}
  \bibinfo{person}{S. Rusinkiewicz}.} \bibinfo{year}{2008}\natexlab{}.
\newblock \showarticletitle{A layered, heterogeneous reflectance model for
  acquiring and rendering human skin}.
\newblock \bibinfo{journal}{\emph{ACM Trans. Graph.}} \bibinfo{volume}{27},
  \bibinfo{number}{5} (\bibinfo{year}{2008}), \bibinfo{pages}{1--12}.
\newblock


\bibitem[\protect\citeauthoryear{Dyer}{Dyer}{1997}]%
        {Dyer97}
\bibfield{author}{\bibinfo{person}{R. Dyer}.} \bibinfo{year}{1997}\natexlab{}.
\newblock \bibinfo{booktitle}{\emph{White}}.
\newblock \bibinfo{publisher}{Routledge}.
\newblock


\bibitem[\protect\citeauthoryear{Fitzpatrick}{Fitzpatrick}{1975}]%
        {fitzpatrick1975}
\bibfield{author}{\bibinfo{person}{T. Fitzpatrick}.}
  \bibinfo{year}{1975}\natexlab{}.
\newblock \showarticletitle{Soleil et peau}.
\newblock \bibinfo{journal}{\emph{J Med Esthet}}  \bibinfo{volume}{2}
  (\bibinfo{year}{1975}), \bibinfo{pages}{33--34}.
\newblock


\bibitem[\protect\citeauthoryear{Friedman, Kahn, and Borning}{Friedman
  et~al\mbox{.}}{2008}]%
        {friedman2008value}
\bibfield{author}{\bibinfo{person}{B. Friedman}, \bibinfo{person}{P. Kahn},
  {and} \bibinfo{person}{A. Borning}.} \bibinfo{year}{2008}\natexlab{}.
\newblock \showarticletitle{Value sensitive design and information systems}.
\newblock \bibinfo{journal}{\emph{The handbook of information and computer
  ethics}} (\bibinfo{year}{2008}), \bibinfo{pages}{69--101}.
\newblock


\bibitem[\protect\citeauthoryear{Kendi}{Kendi}{2019}]%
        {kendi2019}
\bibfield{author}{\bibinfo{person}{I. Kendi}.} \bibinfo{year}{2019}\natexlab{}.
\newblock \bibinfo{booktitle}{\emph{How to be an antiracist}}.
\newblock \bibinfo{publisher}{One world}.
\newblock


\bibitem[\protect\citeauthoryear{Khungurn and Marschner}{Khungurn and
  Marschner}{2017}]%
        {Khungurn17}
\bibfield{author}{\bibinfo{person}{P. Khungurn} {and} \bibinfo{person}{S.
  Marschner}.} \bibinfo{year}{2017}\natexlab{}.
\newblock \showarticletitle{Azimuthal Scattering from Elliptical Hair Fibers}.
\newblock \bibinfo{journal}{\emph{ACM Trans. Graph.}} \bibinfo{volume}{36},
  \bibinfo{number}{2}, Article \bibinfo{articleno}{13} (\bibinfo{date}{April}
  \bibinfo{year}{2017}), \bibinfo{numpages}{23}~pages.
\newblock


\bibitem[\protect\citeauthoryear{Kim}{Kim}{2020}]%
        {KimSciAm20}
\bibfield{author}{\bibinfo{person}{T. Kim}.} \bibinfo{year}{2020}\natexlab{}.
\newblock \showarticletitle{The Racist Legacy of Computer-Generated Humans}.
\newblock \bibinfo{journal}{\emph{Scientific American}} (\bibinfo{year}{2020}).
\newblock
\urldef\tempurl%
\url{https://bit.ly/3thBeye}
\showURL{%
\tempurl}


\bibitem[\protect\citeauthoryear{Phillips}{Phillips}{2020}]%
        {Phillips20}
\bibfield{author}{\bibinfo{person}{A. Phillips}.}
  \bibinfo{year}{2020}\natexlab{}.
\newblock \bibinfo{booktitle}{\emph{Gamer Trouble}}.
\newblock \bibinfo{publisher}{NYU Press}.
\newblock


\bibitem[\protect\citeauthoryear{Sjoding, Dickson, Iwashyna, Gay, and
  Valley}{Sjoding et~al\mbox{.}}{2020}]%
        {Sjoding20}
\bibfield{author}{\bibinfo{person}{M.W. Sjoding}, \bibinfo{person}{R.P.
  Dickson}, \bibinfo{person}{T.J. Iwashyna}, \bibinfo{person}{S.E. Gay}, {and}
  \bibinfo{person}{T.S. Valley}.} \bibinfo{year}{2020}\natexlab{}.
\newblock \showarticletitle{Racial Bias in Pulse Oximetry Measurement}.
\newblock \bibinfo{journal}{\emph{NEJM}} \bibinfo{volume}{383},
  \bibinfo{number}{25} (\bibinfo{year}{2020}), \bibinfo{pages}{2477--2478}.
\newblock


\bibitem[\protect\citeauthoryear{Weyrich, Matusik, Pfister, Bickel, Donner, Tu,
  McAndless, Lee, Ngan, Jensen, and Gross}{Weyrich et~al\mbox{.}}{2006}]%
        {Weyrich2006}
\bibfield{author}{\bibinfo{person}{T. Weyrich}, \bibinfo{person}{W. Matusik},
  \bibinfo{person}{H. Pfister}, \bibinfo{person}{B. Bickel},
  \bibinfo{person}{C. Donner}, \bibinfo{person}{C. Tu}, \bibinfo{person}{J.
  McAndless}, \bibinfo{person}{J. Lee}, \bibinfo{person}{A. Ngan},
  \bibinfo{person}{H.W. Jensen}, {and} \bibinfo{person}{M. Gross}.}
  \bibinfo{year}{2006}\natexlab{}.
\newblock \showarticletitle{Analysis of Human Faces Using a Measurement-Based
  Skin Reflectance Model}.
\newblock \bibinfo{journal}{\emph{ACM Trans. Graph.}} \bibinfo{volume}{25},
  \bibinfo{number}{3} (\bibinfo{date}{July} \bibinfo{year}{2006}),
  \bibinfo{pages}{1013–1024}.
\newblock


\end{thebibliography}

\end{document}